\documentstyle[twoside,fleqn,espcrc2,psfig]{article}
\newcommand{\boneone}{{\beta_{1\times 1}}}
\newcommand{\bonetwo}{{\beta_{1\times 2}}}

\title{ Scaling Analysis of Improved Actions for Pure SU(3) Gauge Theory 
\thanks{Talks by T. Umeda at LATTICE98}}
\author{QCD-TARO :\mbox{ } Ph. de Forcrand \address{SCSC, ETH-Z\"urich, 
                               CH-8092 Z\"urich, Switzerland} ,
        M. Garc{\'\i}a Per\'ez\address{Dept. F\'{\i}sica Te\'orica,
                      Univ. Aut\'onoma de Madrid, E-28049 Madrid, Spain} , 
        T. Hashimoto \address{Dept. of Appl. Phys., Fac. of Engineering,
                           Fukui Univ., Fukui 910-8507, Japan} ,
        S. Hioki \address{Dept. of Physics, Tezukayama Univ.,
                                                Nara 631-8501, Japan} ,
        H. Matsufuru\address{Dept. of Physics, Hiroshima Univ.,
                       Higashi-Hiroshima 739-8526, Japan}$^{\rm ,f}$ ,
        O. Miyamura$^{\rm e}$ ,
        A. Nakamura \address{Res.  Inst. for Inform. Sci. and Education, 
                      Hiroshima Univ., Higashi-Hiroshima  739-8521,
                               Japan } ,
        I.-O Stamatescu$^{\rm f,}$\address{FEST, Schmeilweg 5, D-69118 
                                                Heidelberg, Germany} ,
        T. Takaishi \address{Hiroshima University of Economics,
                                            Hiroshima 731-01, Japan} 
        and
        T. Umeda$^{\rm e}$
          }

\begin{document}
\begin{abstract}

We have explored the behaviour of some improved actions based on
a  nonperturbative renormalization group (RG) analysis in coupling space.
We calculate the RG flow in two-coupling space $(\boneone,\bonetwo)$ and
 examine the restoration of rotational invariance and the
 scaling of physical quantities $( T_c/\sqrt{\sigma})$. 

\end{abstract}
\maketitle
\section{Introduction}
Many improved actions have been proposed and studied which may allow
us much more reliable simulations near the continuum limit within the limited
computer resources existing nowadays.
One way to obtain such improved actions is to find the renormalized
trajectory (RT), any point on which serves as a ``perfect 
action'' \cite{Wilson,iwasaki,perfect}.
Here we will consider the simplest case, i.e., actions in two
coupling $(\boneone,\bonetwo)$ space, which for practical purposes 
are preferable than actions with a larger number of operators.
In this contribution, we evaluate the RG flow and study whether the performance of
such actions improves when they approach the renormalized trajectory.

\section{Search of the Renormalized Trajectory}
\subsection{MCRG}
We adapt Swendsen's blocking transformation scheme \cite{swendBT}, which
is defined as 
\begin{eqnarray}
&&\hspace{-6mm}P(U)=U_{\mu}(x) U_{\mu}(x+{\mu})\\
&&\hspace{-8mm}+\frac{1}{2} \sum_{\nu \ne \mu}U_{\nu}(x)U_{\mu}(x+{\nu}) 
U_{\mu}(x+{\nu}+{\mu})U_{\nu}^{\dag}(x+2{\mu}). \nonumber
\end{eqnarray}

Under this transformation the lattice spacing is doubled. Repeating this 
blocking transformation, we can get an action near the RT.

After each transformation we generate a blocked configuration, the
couplings on which can be determined via the
Schwinger-Dyson equation:
\vspace{-2.5mm}{
\begin{eqnarray}
 \frac{8}{3}\mbox{Re}\langle \mbox{Tr}(U_lG_l^\alpha)\rangle&
\hspace{-4mm}=
&\hspace{-4mm}\sum_\gamma \frac{\beta_{\gamma}}{6}\{-\mbox{Re}\langle 
\mbox{Tr}
(U_lG_l^\alpha U_lG_l^\gamma)\rangle\label{SD}\\
&&\hspace{-33mm}+\mbox{Re}\langle \mbox{Tr}(G_l^\alpha(G_l^\gamma)^\dagger)
\rangle-\frac{2}{3}\langle \mbox{ImTr}(U_lG_l^\alpha)
\mbox{ImTr}(U_lG_l^\gamma)\rangle\}\nonumber
\end{eqnarray}}
\noindent
where $G_l^\gamma$ are the ``staples'' corresponding to the link $U_l$.
Solving these equations, we can calculate the coefficients of operators
corresponding to  the blocked action.
In this paper we will neglect truncation errors and assume that the
resulting blocked actions also have only two
couplings, $\boneone$ and $\bonetwo$.  

\begin{figure}[htb]
\vspace{9pt}
\psfig{file=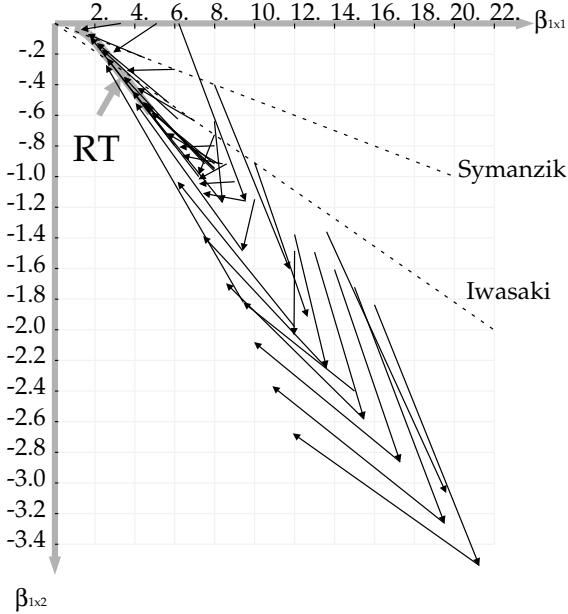,width=75mm}
%\epsfile{file=flow98_proc.ps,scale=0.45}
\caption{Blocking transformation flow in 2 coupling space}
\label{fig.flow}
\end{figure}

Figure~\ref{fig.flow} shows the blocking transformation flow in two-coupling space. 
\vspace{-10mm}
\subsection{Strong coupling limit}
The RT can be calculated within the strong coupling expansion. 
The string tension is 
obtained from the expectation value of Wilson loops.  
To leading order,
\begin{equation} 
\sigma a^2 \simeq -\ln\frac{\langle W_{1\times 2}\rangle}{\langle W_{1\times 1}\rangle} 
\simeq -\ln \left(\frac{\boneone}{18}+\frac{\bonetwo}{\boneone}\right)
\end{equation}
Imposing that in the strong coupling limit,the lattice spacing goes to infinity,
one easily derives the following relation 
\begin{equation}
\frac{\boneone}{18}+\frac{\bonetwo}{\boneone}\rightarrow 0
\mbox{   with   } a \rightarrow \infty
\end{equation} 
As this equation tells us, in this limit, the
RT is a parabola which is presented 
in Figure~\ref{fig.flow} to compare with the RT obtained by solving the S-D
equations as indicated above.

One could as well solve the S-D equations with coefficients calculated by the
strong coupling expansion.  This is work in progress and the results 
will be reported elsewhere.

\section{Scaling analysis}
Here we report on the scaling analysis of several improved actions.
We study the violation of rotational symmetry as well as 
 $T_c/\sqrt{\sigma}$.

\subsection{Violation of rotational symmetry}
\begin{figure}[htb]
\vspace{9pt}
\psfig{file=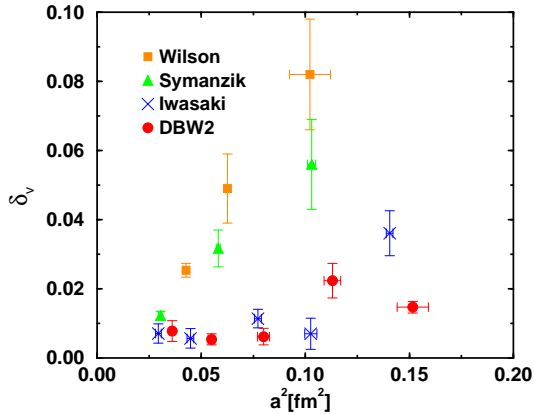,width=70mm}
%\epsfile{file=dV.eps,scale=0.4}
\vspace{-7mm}
\caption{Rotational symmetry violation of various improved actions}
\label{fig.dV}
\vspace{-5mm}
\end{figure}

First we define the following quantity which represents 
the violation of rotational symmetry
\begin{equation} 
\delta^2_V \equiv \sum_{\mbox{off}}\frac{[V(R)-V_{on}(R)]^2}
{V(R)^2\delta V(R)^2}\left(\sum_{\mbox{off}}\frac{1}{\delta V(R)^2}
\right)^{-1}
\end{equation}
where $V(R)$ is the
static quark potential and $\delta V(R)$ its error. $V_{on}(R)$ is
a fitting function from only on-axis data, and $\sum_{\mbox{off}}$ implies
summation over only off-axis data.
We calculate this quantity with various improved actions.
We write $(\boneone,\bonetwo)=\beta (c_{1\times 1},c_{1\times 2})$, and
impose the condition, $c_{1\times 1}+8c_{1\times 2}=1$.
The Wilson action corresponds $c_{1\times 2}=0$. For Symanzik(tree),
Iwasaki and  DBW2 actions, $c_{1\times 2}=-1/12,  
-0.331, -1.4088$, respectively \cite{symact,iwasaki}.
DBW2 (Double Blocked from Wilson action in 2 coupling space)
is defined in ref. \cite{DBW2}.
Results are summarized in Figure~\ref{fig.dV}.

Simulations have been performed on  a lattice of  size  
$12^3 \times 24$ . Thermalization is 5000 sweeps,
while the interval between Wilson loop measurements is 500 sweeps.
We used 100 configurations.

\subsection{$T_c/\sqrt{\sigma}$}
Here we study the scaling behavior of $T_c/\sqrt{\sigma}$ for DBW2. The critical
temperature is defined by
\begin{equation}
T_c=1/N_t a_c \mbox{ : }a_c=a(\beta_c)
\end{equation}
where $N_t$ is the temporal extension of the lattice, 
and $a_c$ is the lattice spacing at critical coupling $\beta_c$.
In order to extract $\beta_c$, we calculate the Polyakov loop susceptibility
on $N_t=4,6$ lattices. Then $\beta_c's$ in the infinite volume 
limit are obtained by using finite size scaling (of $N_t=4$).
The results of $\beta_c$ for DBW2 are presented below.
\\ \\
\begin{tabular}{c|c|c}
\hline \hline
 & $N_t=4$ & $N_t=6$ \\
\hline 
lattice size & $12^3\times 4$ & $18^3\times6$\\
             & $16^3\times 4$ &              \\
\hline
$\beta_c(\infty \mbox{volume})$ & 0.8243(95) & 0.936(25) \\
\hline \hline
\end{tabular}
\\ \\
Next we extract the string tension from the static quark potential at each
value of $\beta_c$.
For DBW2 the results for the static quark potential are  
\\ \\
\begin{tabular}{c|c|c|c}
\hline \hline
$ \beta(N_t) $ & $ A $ & $ \alpha $ & $\sigma $ \\
\hline 
$\beta_c(4)$ & 0.550(17) & -0.255(23) & 0.1555(28)\\
$\beta_c(6)$ & 0.579(96) & -0.357(22) & 0.06996(99)\\
\hline \hline
\end{tabular}
\\ \\ 
where we used the Ansatz 
$V(R)=A+\alpha/R+\sigma R$.

\begin{figure}[htb]
\vspace{9pt}
\psfig{file=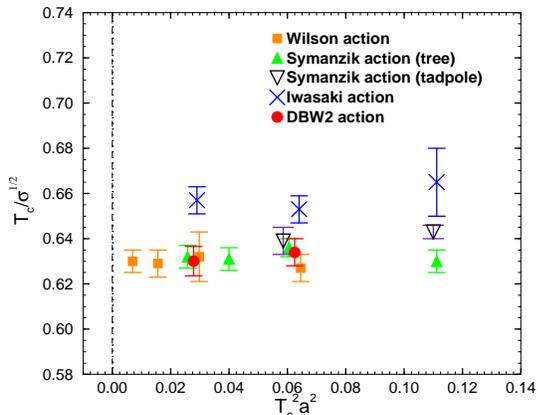,width=70mm}
%\epsfile{file=scaling2.eps,scale=0.45}
\vspace{-12mm}
\caption{Scaling behavior of $T_c/\sqrt{\sigma}$.
We compare DBW2 with results from other actions
in 2-coupling space 
\cite{kaneko,bielefeld}.  }
\label{fig.scale}
\end{figure}

For this action  $T_c/\sqrt{\sigma}$ is $0.6340(60)$ and $0.6301(65)$ at $N_t=4$
and $6$ respectively -  see Fig.~\ref{fig.scale} for a comparison
with other actions in two-coupling space.

\section{Summary}
We have studied the non-perturbative RT of SU(3) gauge theory in two-coupling space.
We find an improvement of rotational invariance as the  improved action
becomes closer to the RT.


\begin{thebibliography}{99}
\bibitem{Wilson} K.G.~Wilson, in {\it Recent Developments in Gauge
theories}, ed. G.~t'Hooft (Plenum Press, New York, 1980) p.363.
\bibitem{iwasaki}Y.~Iwasaki, University of Tsukuba preprint,
                UTHEP-118, 1983.
\bibitem{perfect} P. Hasenfratz, F. Niedermayer, Nucl. Phys. B414 (1994) 
 785.
\bibitem{QCDTARO-MCRG} QCDTARO Collaboration, Phys. Rev. Lett., (1993) 
                        71, 3963.
\bibitem{Okawa} A.~Gonz\'alez-Arroyo and M.~Okawa, Phys. Rev. D35 
 (1987) 672; Phys. Rev. B35 (1987) 2108.   
\bibitem{swendBT} R.H.~Swendsen, Phys. Rev. Lett. 42 (1979) 859.
\bibitem{symact}K.~Symanzik, Nucl. Phys. B226 (1983) 187, 205
\bibitem{DBW2} T.~Takaishi, Phys. Rev. D54 (1996) 1050.
%\bibitem{DBW2} QCDTARO Collaboration, hep-lat/9806008.
\bibitem{kaneko}Y.~Iwasaki and K.~Kanaya, K.~Kaneko, T.Yoshi\'e,
 Phys.Rev.D 56 (1997) 151.
\bibitem{bielefeld}B.~Beinlich, F.~Karsch, E.~Laermann and A.~Peikert
                 hep-lat/9707023
%\bibitem{wilson}G.~Boyd, J.~Engels, F.~Karsch, E.~Laermann, C.~Legeland,
%                M.~L\"utgemeier and B.~Petersson, Nucl. Phys. B469 (1996) 419.
%\bibitem{symanz}F.~Karsch with B.~Beinlich, J.~Engels, R.~Joswig, E.~Laermann,
%                A.~Peikert and B.~Petersson, Nucl.Phys.B (Proc.Suppl.)
%                53 (1997) 413.
\end{thebibliography}
\end{document}